\documentclass[reprint,NumberedRefs]{JASA}

\usepackage{graphicx}
\usepackage{dcolumn}
\usepackage{bm}
\usepackage{color,float}
\usepackage{mathtools}

\usepackage[utf8]{inputenc}
\usepackage[T1]{fontenc}
\usepackage{mathptmx}
\usepackage{etoolbox}

\def\eeq{\relax}
\def\beq#1#2\eeq{\begin{equation}\label{#1}#2\end{equation}}
\def\bal#1#2\eal{\begin{align}\label{#1}#2\end{align}}
\def\bse#1#2\ese{\begin{subequations}\label{#1}#2\end{subequations}}
\def\Im{\text {Im}}

\def\d{\text d}
\def\e{\text e}
\def\i{\text i}

\def\rev#1{{#1}}	

\begin{document}

\title[Constant intensity acoustics]{Constant intensity acoustic propagation 
in the presence of non-uniform  properties and impedance discontinuities: Hermitian and non-Hermitian solutions}

\author{Andrew N. Norris}

\affiliation{Mechanical and Aerospace Engineering, Rutgers University, Piscataway, NJ 08854}%
 \email{norris@rutgers.edu}


\date{\today}

\begin{abstract}
Propagation of sound through a non-uniform medium without  scattering is possible, in principle,  if the density and acoustic compressibility assume complex values, requiring passive and active    mechanisms, also known as Hermitian and non-Hermitian solutions, respectively. Two types of constant intensity wave conditions  are identified: in  the first  the propagating acoustic pressure has constant amplitude, while in the second    the energy flux remains constant.  The fundamental problem of transmission across an impedance discontinuity without reflection or energy loss is solved using a combination of monopole and dipole resonators in parallel.  The solution depends on an arbitrary phase angle which can be chosen to give a unique acoustic metamaterial with both resonators undamped and passive, requiring purely Hermitian acoustic elements. For other phase angles one of the two elements must be active and the other passive, resulting in  a gain/loss non-Hermitian system.  These results prove that \rev{uni-directional and reciprocal} transmission through a slab separating two half spaces is possible using \rev{passive  Hermitian acoustic elements without the need to resort to active gain/loss energetic mechanisms}.  
     
\end{abstract}


\maketitle

\section{Introduction}\label{sec1}

Sound  in non-uniform materials is dominated by multiple scattering  with the result that energy cannot propagate coherently in a given direction without loss.  As a simple example, an acoustic impedance discontinuity  in a one dimensional (1D) acoustic system  causes reflection and resultant loss of energy in the transmitted wave.   Even a single impedance discontinuity can drastically reduce the transmitted energy, as for instance, at an air-water interface.  This effect is reciprocal in  that the same fraction of energy is lost regardless of the incidence direction.   Scattering and loss of forward propagating energy  are inevitable in a classical  or passive, disordered wave medium, now often referred to as a Hermitian system.  


Allowing the acoustic medium to become non-Hermitian, in particular through the introduction of active materials which require external energy input,  makes it possible to avoid  scattering and achieve so-called constant intensity  waves.  This has been demonstrated theoretically and in  simulations   \cite{Makris2015,Makris2017,Brandstoetter2019,Komis2020}.  The application of these ideas to acoustics  was experimentally demonstrated in a 1D waveguide containing scatterers \cite{Rivet2018}.  By adding discrete non-Hermitian elements comprising loudspeakers the backscatter was eliminated and perfect transmission obtained.  
Similar introduction of non-Hermitian elements in 
a multilayered 1D acoustic medium  \cite{An2021} has been shown computationally to achieve constant intensity  wave propagation with no backscatter. 

Non-Hermitian acoustics has been of interest at least since the concept of a  system  with balanced gain and loss, also known as a parity-time (PT) symmetric system,   was  proposed \cite{Zhu2014a} in 2014.   Applications include a metamaterial device formed by a pair of electro-acoustic resonators realizing an acoustic  coherent perfect absorber  \cite{Fleury2015a}, the emergence of exceptional points (EPs) in PT acoustics  \cite{Shi2016},  and coherent acoustic propagation in turbulent fluid flow  \cite{Auregan2017}.  A recent review  \cite{Gu2021} provides an overview of these and other  implications and applications of non-Hermitian acoustics.  \rev{While certainly of interest, non-Hermitian acoustic systems require complicated energy intense active elements, and are to be avoided if purely passive solutions can achieve the same results.}

Our objective is to examine the theory underlying constant intensity acoustic propagation with no backscatter in  disordered systems.  Two types of constant intensity acoustics in one dimension are identified: (i) propagation with constant magnitude of the acoustic pressure, and (ii) propagation with constant   acoustic energy flux. 
Two similar but different equations relating the material proparties are obtained depending on which   constant intensity definition is considered.    We then  focus on constant intensity propagation across a material discontinuity.  We show that  sub-wavelength   actuators, either passive or active, can  ensure perfect transmission with no reflection or energy loss at material discontinuities that otherwise generate reflection and reduce the transmitted energy. 
The point actuators are modeled as acoustic lumped elements, specifically monopole and dipole
 resonators, with an example given using Helmholtz and membrane resonators. 

The  outcome of the  analysis  is a proposal for a new type of metamaterial that provides total  transmission with no loss in energy across an  impedance discontinuity. The metamaterial, or meta-atom since it is a sub-wavelength lumped element, is a {\it passive} undamped combination of monopole and dipole resonators in parallel.  
This demonstrates that  constant intensity acoustic propagation in  a discretely disordered system does not require non-Hermitian elements but may be achieved with the use of standard lossless  passive lumped elements acting in parallel. 

The outline of the paper is as follows.  Basic equations of constant intensity acoustics are derived in Section \ref{sec2}, where the two definitions of constant intensity are distinguished.  The acoustic elements necessary for constant intensity propagation across discontinuities are introduced in Section \ref{sec3}.  The main result of the paper is  in Section \ref{sec4} where it is shown that a combination of passive undamped 
monopole and dipole 
resonators  achieve perfect transmission across an impedance discontinuity with no scattering or energy loss.  Discussion and conclusions are provided in Section \ref{sec5}.


\section{Constant intensity equations}\label{sec2}

\subsection{Basic equations}\label{sec2.1}

Time harmonic motion of frequency $\omega$ is considered with the time-dependent factor $\e^{-\i \omega t}$ omitted. The acoustic momentum and constitutive equations    are 
\bse{1}
\bal{1a}
-\i \omega \rho {\bf v} &= -\nabla p, 
\\
-\i \omega C p&= -\nabla\cdot {\bf v}, \label{1b}
\eal
\ese
where   $p $, $ {\bf v}$, $\rho$, and $C$ are the acoustic pressure, particle velocity,  density, and compressibility (inverse of bulk modulus), respectively, all possibly functions of position ${\bf x}$. Eliminating $ {\bf v}$ yields an  equation for the pressure only
\beq{2}
\nabla\cdot \big(\rho^{-1} \nabla p \big)+\omega^2 C p = 0.
\eeq 
 The density and compressibility may be complex valued, corresponding to a passive system if  $\Im \, \rho \ge 0$ and 
$\Im \, C \ge 0$.   The system is {\it active}, or non-Hermitian, if $\Im \, \rho < 0$ and/or 
$\Im \, C < 0$, requiring an external source of energy to maintain the steady state.  
For the remainder of the paper we specialize to uni-dimensional acoustics ${\bf x} \to x$, so that 
Eq.\ \eqref{2} is 
\beq{2.5}
\big( \rho^{-1}   p' \big)' +\omega^2 C p = 0 
\eeq 
where $p' = \d p/\d x$.

We consider  $\rho$ and $C$ to be functions of position, which, as we will see in Section \ref{sec4}, is critical in dealing with impedance discontinuities.  We also note that the fundamental paper in the field of constant intensity acoustics  \cite{Rivet2018}   assumes constant $C$ with variable density $\rho (x)$, but  uses  the 
incorrect equation $p'' +\omega^2 \rho C_0 p = 0$ 
(Eq.\ (4) of   \citet{Rivet2018}).  

In order to obtain constant intensity uni-directional wave motion we start with the {\it ansatz}
\beq{3}
p(x) = p_0 \, \e^{\i \omega \int_0^x s(x')\d x'}
\eeq
for constant $p_0$, where $s(x)$ has dimensions of slowness. Equation \eqref{2.5} becomes
\beq{4}
s'  -\frac{\rho '}{\rho}  s +      \i \omega  \big(  s^2
- \rho C \big) =0.
\eeq

If both the density and compressibility are constant, Eq.\ \eqref{4} has the solution 
\beq{5}
s(x) = \sqrt{\rho C} = \text{constant}, 
\eeq
as expected for  uni-directional non-dispersive wave propagation. Otherwise $s$ is a function of $x$, and $s\ne \sqrt{\rho C}$. 

According to Eqs. \eqref{1a} and \eqref{3} the particle velocity 
is 
\beq{1-1-1}
v = \frac s \rho p. 
\eeq
The ratio $v/p$ is therefore the local acoustic admittance
 $y = s/\rho$, or equivalently, we can identify $p/v$ as the local impedance 
$z =  \rho /s$ $(=1/y)$. The acoustic impedance is $z=\rho c$ where 
$c=1/s$ is the local phase speed.
Equation \eqref{4} can be rewritten as an equation for the impedance
\beq{8}
z'+    \i \omega  \big( C z^2 -\rho  \big) =0
\eeq
which shows that in regions where $\rho (x)/C(x) =$ constant that 
the impedance $z$ (equivalently the admittance $y$) is constant 
and given by $z=\sqrt{\rho/C}$.  

The Riccati equations \ \eqref{4} and \eqref{8}  provide  relations for the possibly complex-valued  density and/or compressibility in terms of the variation in phase derivative $s$ and impedance $z$, respectively.  We require that $s$ remains real, so  setting $\rho = \rho_1 +\i \rho_2$ and $C=C_1+\i C_2$ the real and imaginary parts of 
Eq.\ \eqref{4} imply
\bse{2=3}
\bal{23a}
s'+\omega(\rho_1 C_2 + \rho_2 C_1) - s (\log |\rho|)' &=0,
\\
s^2 -\rho_1 C_1 +\rho_2 C_2 +s\frac {\rho_1\rho_2}{\omega |\rho|^2} 
(\log \rho_1/\rho_2 )' 
&=0.
\eal
\ese

\subsection{Example: Total energy absorption}

Assuming  $\rho$ to be constant and real,  Eq.\ \eqref{2=3} gives 
\beq{0-2}
C = \frac 1 \rho \big(s^2 -\i \omega^{-1}s'\big). 
\eeq
This complex-valued compressibility is passive or active depending as $s'\le 0$ or $s' > 0$, respectively.  The physical consequences  can be seen through the energy flux averaged over a period, 
\beq{0-4}
F(x) = s(x)\frac {|p_0|^2} {2\rho} . 
\eeq
Energy flux is constant if $s$ is also  constant. 
A decreasing $s$ as a function of $x$ leads to a diminishing energy flux, consistent with the passive damping caused by $\Im\, C >0$. Conversely, $F$ increases with $x$ if $s' > 0$, corresponding to an active non-Hermitian acoustic compressibility. 

Consider the following example for $s(x)$ with associated  passive compressibility with damping 
\beq{2=5}
s= - \frac x a s_0, \quad
\Rightarrow \quad C = C_0\Big(
\frac{x^2}{a^2} + \frac \i {\omega s_0 a}
\Big)
\eeq
 where $C_0 = s_0^2/\rho$.   According to Eq.\ \eqref{0-4} the energy flux on either side of $x=0$ is directed towards $x=0$ with $F(0) = 0$.  The compressibility of Eq.\ \eqref{2=5}, which is a symmetric function of $x$, causes waves from either direction to be totally absorbed, with no backscatter.  The material acts as an acoustic black hole. 

\subsection{Constant amplitude or constant energy?}

The above example shows that even though the pressure amplitude remains constant the associated wave energy can decrease or increase. This suggests a more reasonable definition of "constant intensity"  that the energy flux is constant even as the impedance changes.  With that in mind the following  {\it ansatz} is more appropriate than Eq.\ \eqref{3}:
\beq{38}
p(x) = p_0 \, \sqrt{\frac{z}{z_0}}\,   \e^{\i \omega \int_0^x s(x')\d x'}
\eeq
for constant $p_0$,  $z_0$ where from now on the impedance is defined  $z(x) \equiv \sqrt{\rho (x)/C(x)}$.  This solution provides the same energy flux in regions of different but constant  $z$.   \rev{A Riccati equation analogous to Eq.\ \eqref{8} follows from Eq.\ \eqref{38} 
\beq{41}
q'+    \i \omega  \big( \rho q^2 - C  \big) =0
\eeq
where $q = s/\rho + z' /(2\i \omega \rho z)$.} We do not pursue this equation further.  We will however provide an exact solution  to Eq.\ \eqref{38} in the next Section for the important case of a point discontinuity in acoustic properties. 

Our  focus on a point discontinuity in impedance as compared with a continuous variation is because the latter  requires a continuous variation in possibly active material properties. Any discontinuity in impedance leads to scattering (reflection) and loss of transmitted energy.  We therefore consider what is necessary to maintain the forward propagating wave at constant intensity according to Eq.\ \eqref{38} in the presence of point  changes in impedance.  In order to address this we will take a step back to the fundamental equations \eqref{1} since  Eq.\   \eqref{41} 
is based on smoothly varying impedance and is not suitable for point discontinuities. 
First, we introduce the elements to achieve one-way constant intensity propagation across the discontinuities.

\section{Monopole and dipole resonators}\label{sec3}

Monopole and dipole resonators have been the workhorses of acoustic metamaterials: Helmholtz resonators (monopoles) provide the basis for negative effective compressibility   \cite{Fang06,LeeParkSeoEtAl2009,LeeParkSeoEtAl2010,Lai2011,Lee2016}, while dipole membrane resonators yield negative effective inertia  \cite{Lee2009,LeeParkSeoEtAl2010,SeoParkLeeEtAl2012,Lee2016}.  Together they lead to the possibility of so-called doubly negative properties in acoustic metamaterials; in fact the earliest notions in metamaterials are generally agreed to revolve around concepts of negative properties, see for instance the historical review of  \citet{Carlin1964}.  Monopole and dipole resonators in combination can provide total energy absorption  \cite{Yang2017}, using  a design that is closely related to the present one proposed in the next Section  \cite{Yang2015}. Despite the recent successful use of sub-wavelength acoustic lumped elements it should be borne in mind that they have a century long history in the development of analog acoustic control, particularly in acoustic wave filters  \cite{Stewart22,Mason1928,Lindsay1929}. 

\begin{figure}[H]
\centering
\includegraphics[width=3.4in]{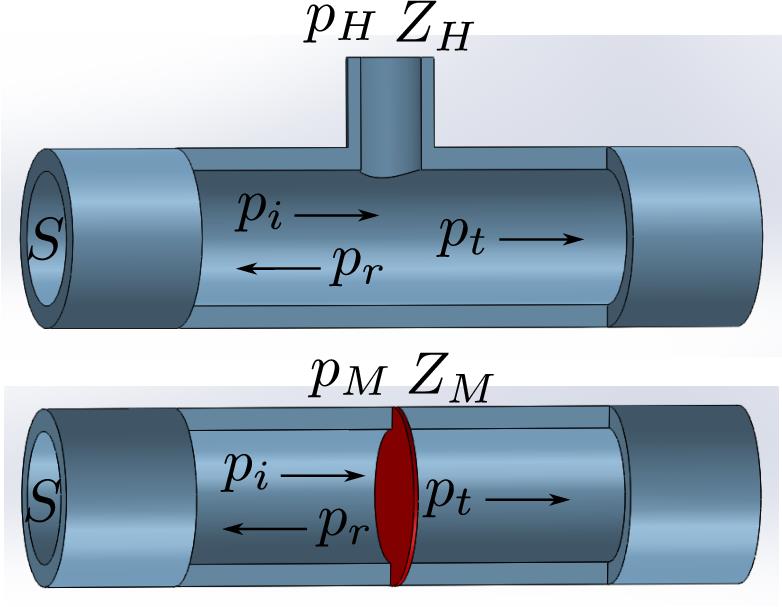}
\\
\caption{Reflection and transmission from a monopole (top) and a dipole membrane resonator, both considered as sub-wavelength  elements  at the interface between different acoustic fluids on the left and right.  \rev{The monopole shown is a Helmholtz resonator, but other types are possible, such as the Hybrid Membrane (HM)  resonator or a pair of Decorated Membrane (DM)  resonators \cite{Yang2015}.}  }
\label{fig1}
\end{figure}

Lumped parameter acoustic elements consisting of a Helmholtz resonator and a sprung mass or membrane are shown in  Figure \ref{fig1}. In preparation for dealing with constant intensity across an interface we first consider each  element situated between acoustic fluids with properties 
$\rho_-$, $C_-$ on the left and $\rho_+$, $C_+$ on the right. The membrane is placed so that it separates the  fluids.  As suggested in the Appendix,  the Helmholtz resonator fluid can be taken as one of the two with the neck inserted in the respective fluid close to the interface.   \rev{Alternatively, the monopole element could be provided by a Hybrid Membrane (HM)  resonator \cite{Yang2015} situated the same way as the HR in Figure \ref{fig1}, or by a pair of Decorated Membrane (DM)  resonators \cite{Yang2015} placed parallel to the membrane resonator separating the fluids as in Fig.\ 1(a) of \cite{Yang2015}.    }

With $p$ as acoustic pressure, $U = Sv$ as volumetric velocity, the 
incident, reflected and transmitted quantities are related by the $U_i = p_i/Z_-$,  $U_r = -p_r/Z_-$,  $U_t = p_t/Z_+$, where $Z=z/S$, with 
$z=\sqrt{\rho/C}$ the acoustic impedance and $S$ the waveguide cross-sectional area, Figure \ref{fig1}.    The  Helmholtz and membrane resonators are defined by  impedances $Z_H$ and $Z_M$ such that  $U_H = p_H/Z_H$ and $U_M = p_M/Z_M$, respectively.  The systems are passive if 
$\Re \, Z_H \ge 0$, $\Re \, Z_M \ge 0$, they are otherwise  active.  
Models for both types of acoustic elements are discussed in the Appendix.  

The conditions for the attached Helmholtz resonator transmission-reflection problem are 
\beq{1-}
\begin{aligned}   
p_i+p_r &= p_t = p_H, \\
U_i+U_r &= U_t+U_H ,
\end{aligned}   
\eeq
while those for the in-line mass-spring system are 
\beq{2-}
\begin{aligned}   
p_i+p_r &= p_t + p_M, \\
U_i+U_r &= U_t= U_M.
\end{aligned}   
\eeq
Solving for the reflection and transmission coefficients, defined by $p_r=Rp_i$, $p_t= Tp_i$, gives
\bse{3-}
\bal{3a}
R_H &= \frac{\frac 1{Z_-}-\frac 1{Z_+}-\frac 1{Z_H}}{\frac 1{Z_-}+\frac 1{Z_+}+\frac 1{Z_H}},  \ \  T_H = 1+R_H ,
\\
R_M &= \frac{Z_M+Z_+ - Z_-}{Z_M+Z_+ +Z_-},  \ \  T_M = \frac{Z_+}{Z_-}\big( 1-R_M \big)  .
\label{3b}
\eal
\ese

If $Z_+ > Z_-$  a transmission coefficient of unity can be achieved with the Helmholtz resonator by choosing $\frac 1{Z_H} = \frac 1{Z_-}-\frac 1{Z_+}$, yielding $R_H=0$ and $T_H = 1$.  This passive but dissipative element achieves one way propagation but at the expense of   energy  loss since a transmission coefficient of magnitude 
$\sqrt{Z_+/Z_-}$ is necessary to maintain the same energy flux in $x>0$ as for the incident wave in $x<0$ (see next section for details). 

Similarly, one way propagation  across the discontinuity with $Z_+ < Z_-$
is achieved with the passive membrane of impedance  $Z_M = Z_- - Z_+$, which gives $R_M=0$ and $T_M = \frac{Z_+}{Z_-}$.  Again, this is less than the required transmission amplitude  $\sqrt{Z_+/Z_-}$ for maintaining the wave energy, with the energy loss associated with the damping effect of the membrane.

\section{Propagation across impedance discontinuities}\label{sec4}

 Here we show that constant intensity propagation through an impedance discontinuity 
can be achieved with two passive undamped elements at the point of discontinuity, or by  a combination of one active element with a damped passive element.  

\begin{figure}[H]
\centering 
\includegraphics[width=3.4in]{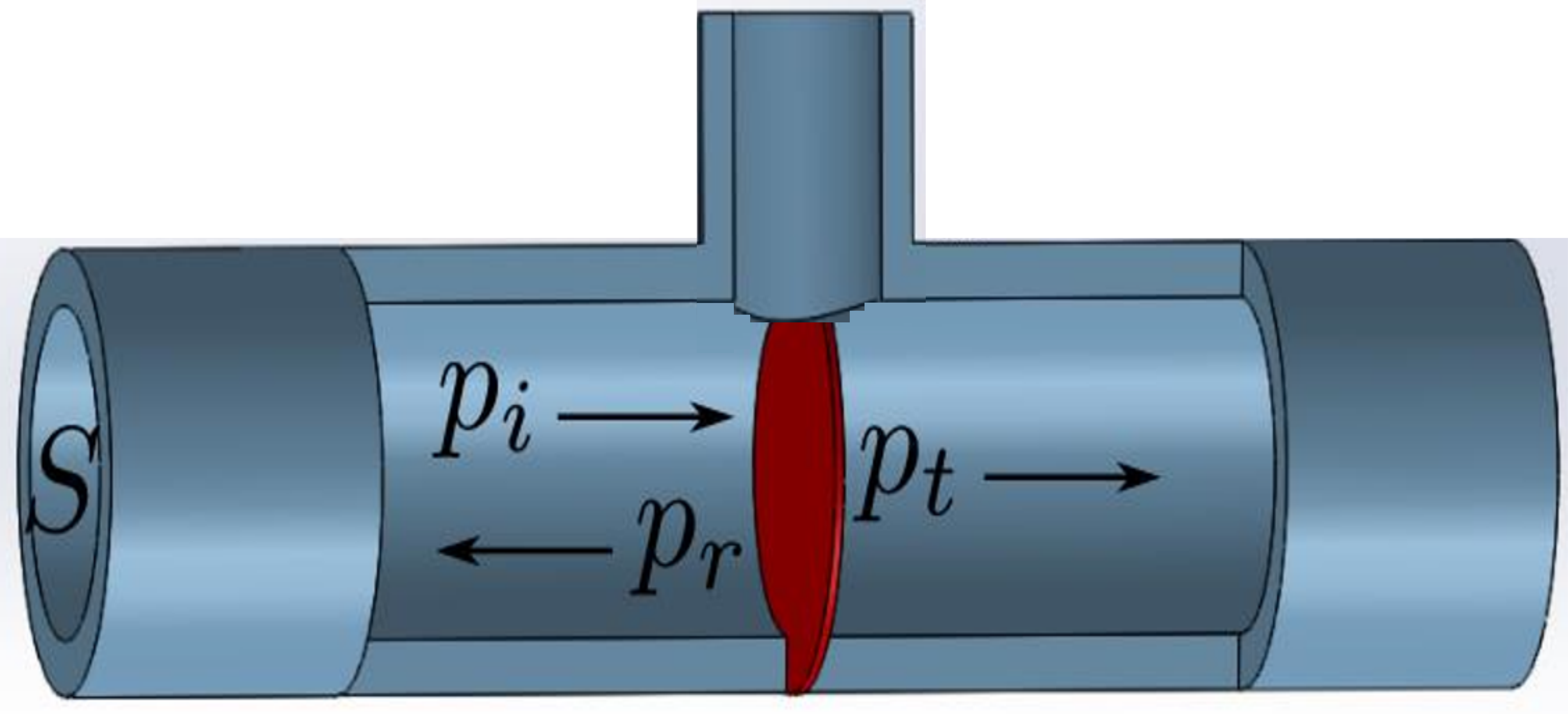} 
\caption{The combined monopole+dipole elements in parallel.  }
\label{fig2}
\end{figure}

\subsubsection{Monopole and dipole strengths}

We consider an acoustic waveguide with an impedance  discontinuity at $x=0$, i.e. 
 $Z(\pm 0) = Z_\pm$. This could be due to different fluids in a waveguide of constant cross-section, or a change in cross-sectional area with the same fluid on either side.  
   In the neighborhood of $x=0$ we therefore have $Z(x) = Z_- + \Delta Z \,H(x)$ where $\Delta Z = Z_+ - Z_-$ and $H$ is the Heaviside step function.  We  consider the {\it ansatz} 
\bse{-33}
\bal{-33a}
\rho(x) &= \rho_0(x) + \i \alpha \,\delta(x), 
\\
C(x) &= C_0(x) + \i \beta \,\delta(x), 
\eal
\ese
where $\rho_0(x) = \rho_- + \Delta \rho \, H(x)$,  $C_0(x)=C_-+ \Delta C \, H(x)$,   $\delta = H'$ is the delta function, and $\alpha$ and $\beta$ are constants. 
The concentrated density and compressibility of Eq.\ \eqref{-33} correspond to a combination of 
dipole and monopole resonators  in parallel,  as exemplified in Fig.\ \ref{fig2}.  A similar parallel combination of monopole and dipole resonators was used  \cite{Yang2015} to achieve a quite different effect: total absorption.  Apart from the  different application, we are here considering the pair of elements positioned right at the discontinuity of impedance, whereas Yang et al.\  \cite{Yang2015} had no discontinuity. 

Equations \eqref{1a} and \eqref{1b} now become 
\bse{1=1}
\bal{1=1a}
\i \omega \big(\rho_0(x) + \i \alpha \,\delta(x)\big)  v &= p', 
\\
\i \omega \big(C_0(x) + \i \beta \,\delta(x) \big) p &= v'. \label{1=1b}
\eal
\ese
Both $p$ and $v$ \rev{are smooth functions of $x$ near the interface but they} can be discontinuous at $x=0$.   \rev{To be specific, we assume they have distinct values on either side of the interface:} $p(0\pm) = p_\pm$ and
$v(0\pm) = v_\pm$.  
\rev{Equations \eqref{1=1} imply jump conditions connecting the values of $p$ and $v$ across the interface.  The functions multiplying the Dirac delta functions in Eqs.\ \eqref{1=1} are themselves discontinuous precisely at $x=0$, and therefore  some care needs to be taken in deriving the jump conditions in the usual manner by integration.   We use the identity
$\int \delta(x) v(x) \d x = \frac 12 (v_+ + v_-)$, which can be seen by noting that the  delta function is symmetric in $x$ and can be written in various forms as a limit.  For instance, taking $\delta(x) = \lim_{\epsilon\to 0}\frac 1{\epsilon} \big(
H(x+\frac{\epsilon}2) - H(x-\frac{\epsilon}2) \big) $ and integrating from $x=-\frac{\epsilon}2$ to $x=\frac{\epsilon}2$ gives the identity.  Note that this identity is not universally applicable but depends upon the context, with some embarrassing results when used where it should not be \cite{Griffiths1999}. }
In summary, Eqs.\ \eqref{1=1} give
\bse{1=2}
\bal{1=2a}
\Delta p  &= -\frac{\omega \alpha}2  (v_+ +v_-), 
\\
\Delta v  &= -\frac{\omega \beta}2 (p_+ +p_-), 
\eal
\ese
where $\Delta p = p_+ - p_-$ and $\Delta v = v_+ - v_-$  
provide the required monopole and dipole strengths.

The connection between  the point scatterers of Eq.\ \eqref{-33} and the membrane and Helmholtz resonators can be found by first considering reflection and transmission from the monopole scatterer, i.e. Eq.\ \eqref{-33} with $\alpha =0$.  In this case $p$ is continuous at $x=0$, and a  simple analysis gives
\beq{14}
Z_H = \frac 1{\omega \beta S}. 
\eeq
Conversely,  reflection and transmission from the dipole  scatterer, i.e. Eq.\ \eqref{-33} with $\beta=0$ has  $v$ continuous at $x=0$ with 
\beq{15}
Z_M = \frac {\omega \alpha} S.
\eeq

For wave incidence from $x<0$, we have 
\beq{1=39}
p(x) = \begin{cases}
e^{\i k_- x} + R e^{-\i k_- x}, & x<0, 
\\
Te^{\i k_+ x} , & x>0, 
\end{cases}
\eeq
where $k_\pm = \omega /c_\pm$.
From Eqs.\ \eqref{1=2} through   \eqref{1=39} we find
\bse{1=4}
\bal{1=4a}
T &=   \frac{ 2Z_+ (Z_H-\frac 14 Z_M)}
{ Z_M Z_H + Z_+ Z_-  +(Z_H +\frac 14 Z_M)(Z_+ +Z_-) },
\\
R &=   \frac{ Z_M Z_H - Z_+ Z_-  +(Z_H +\frac 14 Z_M)(Z_+ -Z_-) }
{ Z_M Z_H + Z_+ Z_-  +(Z_H +\frac 14 Z_M)(Z_+ +Z_-) }.
\eal
\ese

Setting $R=0$ requires that the impedances are related by  
\beq{1=5}
Z_M = \frac{Z_+Z_--Z_H\Delta Z}{ Z_H + \frac 14 \Delta Z} 
\eeq
and the resulting transmission coefficient is 
\beq{1=6}
T = \frac{Z_+}{Z_-} \Big(
\frac{2Z_- -Z_M}{2Z_+ +Z_M}
\Big).
\eeq
Our objective is to find a metamaterial that achieves a designated transmitted wave amplitude.   With that in mind we rewrite the previous equations to 
 express $Z_M$ and $Z_H$ in terms of $T$ \rev{and the impedances on either side of the interface}:
\bse{1-6}
\bal{16a}
Z_M &= \frac{2Z_+Z_- (1-T)}{ Z_+ +Z_-T}, 
\\
Z_H &= \frac{Z_+Z_- (1+T)}{ 2(Z_+ -Z_-T)} .
\eal
\ese

\subsubsection{Explicit expressions for the monopole and dipole impedances}

The incident energy flux, averaged over a cycle, is $\frac 12 S\,\Re \,(pv^*)$ for $x<0$, which is $\frac 12 \Re\, Z_-^{-1}$.  The corresponding transmitted energy flux is $\frac 12 |T|^2 \Re \, Z_+^{-1}$. Assuming real-valued impedances 
$Z_\pm$, this means that in order to maintain the energy flux, or wave energy, we need 
\beq{1-7}
T =  e^{\i \phi}\, \sqrt{\frac{Z_+}{Z_-} } 
\eeq
for arbitrary phase angle $\phi$. Equations \eqref{1-6} become  
\bse{1-8}
\bal{18a}
Z_M &=  \frac{-2\sqrt{Z_+ Z_-}( \Delta Z\cos \phi + \i (Z_+ + Z_-) \sin \phi )} 
{ Z_++Z_- +2 \sqrt{Z_+ Z_-} \cos \phi  } ,
\\
\frac 1{Z_H} &= \frac{ \Delta Z\cos \phi - \i (Z_+ + Z_-) \sin \phi}
{\frac 12 \sqrt{Z_+ Z_-}( Z_++Z_- +2 \sqrt{Z_+ Z_-} \cos \phi ) } ,
\eal
\ese
or after some simplification, 
\bse{1=}
\bal{1=a}
Z_M &=  -\i 2Q\,\sqrt{Z_+ Z_-}\,  \e^{-\i \theta},
\\
{Z_H} &= (-\i 2Q)^{-1}\, \sqrt{Z_+ Z_-}\,\e^{-\i \theta}, 
\eal
\ese
where
\beq{01}
Q = 
\sqrt{
\frac
{ Z_++Z_- -2 \sqrt{Z_+ Z_-} \cos \phi  }
{ Z_++Z_- +2 \sqrt{Z_+ Z_-} \cos \phi  }
 }
 , \ \
\tan\theta = \frac { Z_+ -Z_-   }
{ Z_++Z_- }\cot \phi .
\eeq

The explicit expressions of Eq.\ \eqref{1=}, or the relations  
\beq{1-9}
Z_M = {-}{4Q^2} Z_H, \ \ 
Z_M Z_H^* =Z_M^* Z_H   = - Z_+ Z_-  ,
\eeq
indicate that if one of the resonators is passive and lossy, e.g. Re$\, Z_M  >0$, then the other is active with gain,  Re$\, Z_H  <0$. Since the transmitted acoustic pressure  has the same energy as that of the incident wave and because there is  no reflection, the gain and loss of the two elements must cancel each other in order to conserve wave energy.   

Both elements are passive  if $\cos \phi  =0$, in which case, taking 
$\sin \phi = \mp 1$, we have a unique pair of undamped passive solutions
\beq{1+7}
Z_M = \pm 2 \i  \sqrt{Z_+ Z_-}, 
\quad
Z_H = -\frac{Z_M} 4.
\eeq
The elements are lossless by virtue of the fact that both have zero real part.  Furthermore, if one is mass-like $(\Im\, Z <0)$ the other acts as a stiffness $(\Im\, Z >0)$.    

\subsubsection{Example:  transmission through an air-water interface}

As an example we consider a water-air interface, a  common acoustic boundary but with a rather  extreme impedance mismatch of approximately 3,644 to 1. 
We assume that the interface is equipped with membrane and Helmholtz resonators in parallel  with impedances  given by Eqs.\ \eqref{A1} and \eqref{A2}$_1$. In order to satisfy the  inequality $Z_M^* Z_H    <0$, from Eq. \eqref{1-9},  the target   frequency for full transmission, $\omega= \omega_0$ must lie between the smaller and larger of $\{ \omega_M, \omega_H\}$.  To be specific, we assume $\omega_M < \omega_H $ and therefore $\omega_M < \omega_0 < \omega_H $.  Furthermore, for the sake of simplicity we assume that $S_+=S_- = S$, and so  $Z_\pm = z_\pm/S$.  Equation \eqref{1+7} with $Z_M = - 2 \i  \sqrt{Z_+ Z_-} $
and Eqs.\ \eqref{A1} and \eqref{A2}$_1$ then imply 
\bal{2+5}
\frac{\omega m}{2S}\Big( 1-\frac{\omega_M^2}{\omega^2} \Big)
&= \frac{2S}{\omega C_b V}\Big( 1-\frac{\omega^2}{\omega^2_H} \Big)
\notag \\ &= \sqrt{z_+ z_-}  \quad\text{for} \quad
\omega   = \omega_0 .
\eal

The material impedances $z_\pm$ are defined by the water-air interface,  with $\rho$ and $c$ for water and air as 
$1,000$ kg/m$^3$, $1,500$ m/s and $1.2$ kg/m$^3$, $343$ m/s, respectively. 
We then select  the following parameters: the HR bulb and neck fluids and hence the values of $C_b$ and $\rho_n$, the waveguide cross-sectional area $S$,  the membrane and Helmholtz resonance  frequencies $f_M$,  $f_H$ and the frequency for total transmission $f_0=\omega_0/(2\pi)$.  Equation   \eqref{2+5} then implies values for the membrane mass 
$m$ and Helmholtz resonator volume $V$, and Eq.\ \eqref{A2}$_2$ then yields the ratio $A/l$ for the HR neck (see the Appendix). 

For the example considered, the HR bulb and neck fluids are both  air, the full-transmission frequency is $f_0=170$ Hz, the resonators frequencies are 
$f_M=100$ Hz, $f_H=200$ Hz, and the waveguide cross-section is circular of radius $0.1$ m.  We then find from Eq.\ \eqref{2+5} that 
$m=2.23$ kg and $V=92.7$ cm$^3$, and from Eq.\ \eqref{A2}$_2$ that $A/l = 1.2\,\text{x} 10^{-3}$ m.   The latter corresponds, for instance, to a cylindrical HR neck of radius $0.75$ cm and length $l=14.2$ cm.  
These parameters ensure perfect transmission at the chosen operating frequency, in this case $f_0=170$ Hz. 

\begin{figure}[H]
\centering 
\includegraphics[width=3.4in]{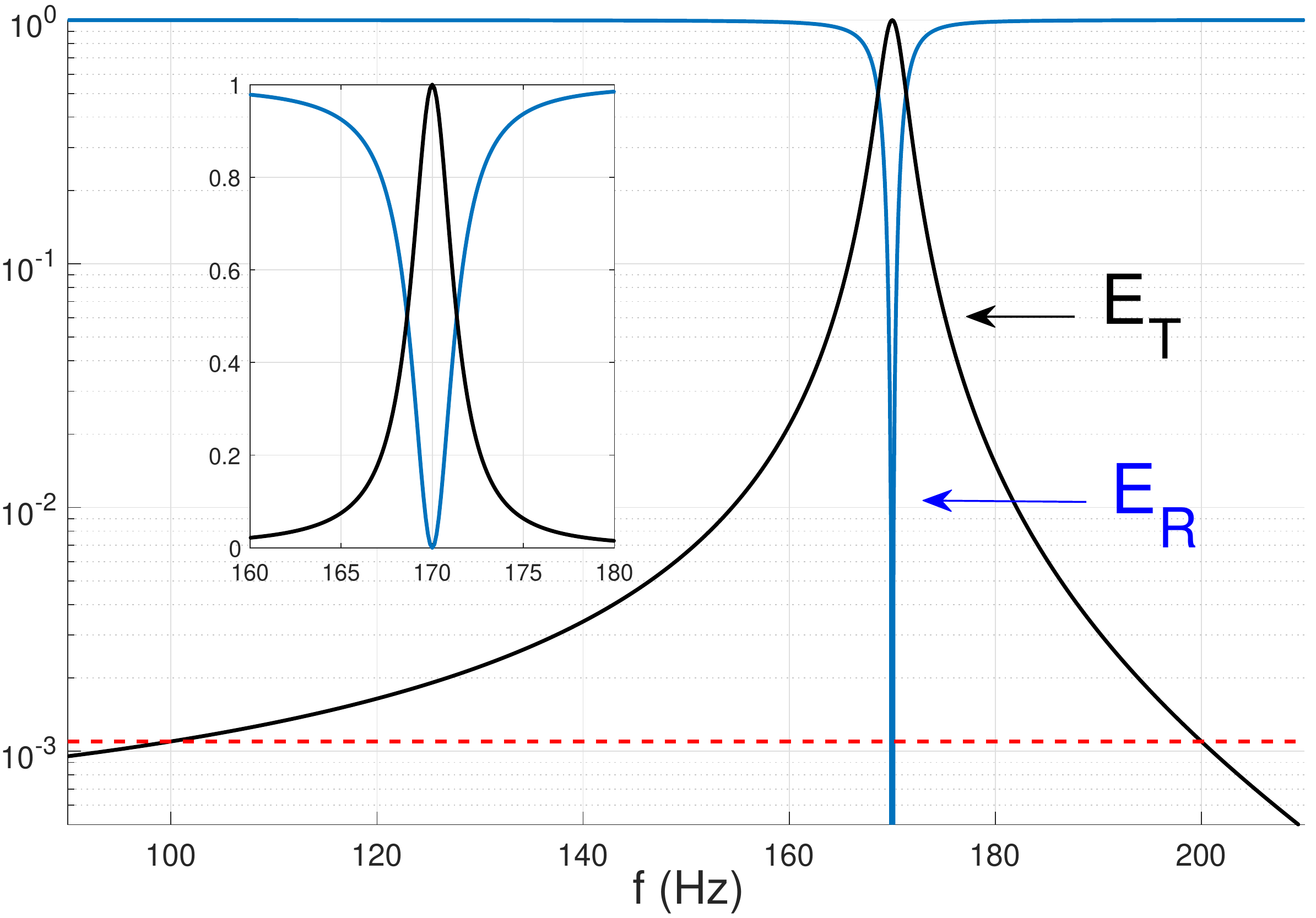} 
\caption{The transmitted and reflected energy fluxes, $E_T$ and $E_R$ of Eq.\ \eqref{66}, as a function of frequency. The dashed curve is the value of $E_T$ in the absence of the passive resonators: $E_T = 1.1\,\text{x} 10^{-3}$.  }
\label{fig3}
\end{figure}
Figure \ref{fig3} shows the  fractional   reflected and transmitted energy fluxes as a function of frequency for these parameters, where  
\beq{66}
E_R = |R|^2, \quad
E_T = \frac{Z_-}{Z_+}|T|^2 ,  
\eeq
with   $E_R+E_T = 1$. The inset in Fig.\ \ref{fig3} indicates perfact transmission at the selected frequency $f_0=170$ Hz.  Significant transmission $(E_T \ge 0.5)$ occurs in a narrow bandwidth of about 4 Hz.  However, $E_T$ is greater than the ambient value $ 1.1\,\text{x} 10^{-3}$ of an air-water interface for every frequency over the broad bandwidth between the two resonance frequencies $f_M $ and $f_H$.  Figure \ref{fig3} shows that the transmitted energy equals the ambient value at 
both $f_M $ and $f_H$, and it lies below the ambient value for frequencies less than $f_M$ and greater than $f_H$. The pair of passive resonators therefore act as a broadband transmission amplification device that achieves full transmission over a narrow range of frequencies centered at $f_0$.

\section{Discussion and conclusions}\label{sec5}

The main result of this paper is that \rev{ zero-scatter one-way propagation    in the presence of an impedance  discontinuity can be achieved by  purely passive methods.  There is no need to resort to  more complicated active systems using gain/loss mechanisms characteristic of  non-Hermitian acoustics.  }

Two ways of looking at  constant intensity acoustics for one dimensional wave motion have been considered: constant pressure amplitude or constant energy flux. Of the two, the constraint of constant energy is more sensible.  This is evident in the solution for transmission across a material discontinuity.  A change in  impedance causes scattering, but introduction of the combined monopole+dipole acoustic "meta-atom" of Eq.\ \eqref{1-8} results in total transmission with no loss in energy.  This solution contains an arbitrary phase $\phi$ which "rotates" the solution in the space of passive/active metamaterials. When $\cos\phi = 0$ the metamaterial  becomes passive and undamped.  Otherwise, the two elements form a gain/loss pair with one  active and the other passive and damped that together ensure  that the transmitted wave has the same energy as  the incident one. 

Transmission through the  monopole+dipole system \eqref{1-8} or \eqref{1+7} is reciprocal in that it provides transmission with full energy conservation and zero reflection for incidence from either side of the impedance discontinuity.  This means, for instance, that perfect transmission through a finite slab or layer of material can be achieved by placing the same monopole+dipole system at each end. 

In conclusion, the paper provides   fundamental results relevant to sound wave amplification and scattering cancellation.  The proposed acoustic metamaterial comprising a pair of tuned resonators, passive/passive or active/passive, could enable a huge increase in acoustic communication efficiency across disparate  interfaces, as shown in the example for passive transmission amplification  between air and water.

\begin{acknowledgments} 
This work was supported by NSF EFRI award no. 1641078. 
\end{acknowledgments}

\appendix* 

\section{Impedance models}\label{appA}

\subsection{Membrane or dipole impedance}

The  impedance $Z_M$  for a lossless membrane resonator follows from e.g.  \citet{Lee2009,Lee2016} as
\beq{A1}
Z_M = \frac{-i\omega m}{S^2}\Big( 1-\frac{\omega_M^2}{\omega^2} \Big) ,  \ \  
\omega^2_M = \frac {\kappa}m 
\eeq
This can be envisaged as a membrane separating the fluid on either side, and subject to a pressure differential across it. The mass $m$ is the total mass, and the stiffness $\kappa$ arises from the elasticity as it is stretched under pressure.  

A more precise model, due to  \citet{Ingard1954} and based on  \citet[p.\ 201]{Morse1948}, considers the dynamics of a membrane of wavenumber $k_M= \omega /c_M$, radius $a$, yielding 
\beq{9-2}
Z_M = \frac {i  \omega m}{ S^2}\frac{J_0(k_Ma)}{J_2(k_Ma)} .
\eeq 
The resonant frequency of the membrane, $\omega_M$, can be calculated analytically by considering free flexural vibration of circular plate clamped all around the circumference. 
The general solution 
in polar coordinates is  \cite{Leissa} 
$W_n=\big (A_nJ_n(k_M r)+C_nI_n(k_M r)\big )\cos n\theta$,
where $k_M^4=\rho h\omega^2/D$; $J_n$ and $I_n$ are the Bessel functions of the first kind and the modified Bessel functions of the first kind, respectively. Clamped boundary conditions 
$W = \frac{\partial W}{\partial r}=0$  at $r=a$
lead to the eigenvalue problem
$J_n(k_M a)I_{n+1}(k_M a)+I_n(k_M a)J_{n+1}(k_M a)=0$.
The natural frequencies of the circular membrane plate can be calculated from the  roots  $k_M a$. Here we are only interested in the fundamental mode, i.e. $(0,1)$ mode. Taking the number of nodal diameter as $n=0$ and only solving for the first mode, we have $k_M a=3.1962206\cdots$.
Among related models for the acoustic impedance  of internal mass-spring resonators   \cite{Liu000,Mace14} and  membrane oscillators we note the explicit expression of    \citet{BongardLissekMosig2010}:
\beq{effimp}
Z_M=\frac{-i\omega m}{S^2}\frac{J_1(k_M a)I_0(k_M a)+I_1(k_M a)J_0(k_M a)}{J_1(k_M a)I_2(k_M a)-I_1(k_M a)J_2(k_M a)}.
\eeq

\subsection{Helmholtz resonator or monopole impedance}

The Helmholtz resonator (HR) impedance is a classic result   \cite{Stewart1925,Stewart1926a}   usually considered in the context of a uniform ambient acoustic medium.  The HR comprises a volume of fluid - the bulb- which acts as a mechanical spring, and a volume of fluid - the neck - acting as a mass. The present application envisages the Helmholtz resonator located at an impedance discontinuity, Fig.\ \ref{fig2}.  The placement of the neck can therefore be in either of the fluids, but is perhaps simpler if the fluid in the neck is the same as that into which the neck enters. In the example considered in \S \ref{sec4} the fluid is air.  The classical HR is normally considered to contain a single acoustic  fluid but here we consider the possibility of different fluids in the bulb and neck.  In order to avoid mixing of the two acoustic fluids  it would be  necessary to introduce an impermeable membrane with ignorable acoustic influence.  That means a material both light and compliant enough that it does not cause pressure or velocity jumps, while separating the acoustic fluids. The reason we consider the possibility of two fluids is to increase the potential range of the HR parameters, which for the purposes considered here can be very demanding.  At the same time, the fluids should be such that the potential and kinetic energies over one cycle are predominantly located in the bulb and neck fluids, respectively.

With these considerations in mind, generalizing  \S10 of   \citet{KinslerFrey} to the two distinct fluids in the bulb and neck, we model the 
undamped HR  as
\beq{A02}
Z_H = \frac{1}{-\i\omega C_b V}
-\i\omega \rho_n \frac lA 
\eeq
where $V$ is the volume of the bulb acoustic fluid with compressibility $C_b$, and $\rho_n$, $A$, $l$ are the neck fluid density, cross-sectional area and length, respectively.  The  impedance is therefore
\beq{A2}
Z_H = \frac{1}{-\i\omega C_b V}\Big( 1-\frac{\omega^2}{\omega^2_H} \Big) ,  \ \  \omega^2_H = \frac{A}{\rho_nC_b Vl}
\eeq
where the resonance frequency $\omega_H$ depends upon the bulb compressibility $C_b$ and the neck fluid density $\rho_n$, along with the geometric parameters $V$, $A$ and $l$.


\end{document}